\documentclass[twocolumn,prb,superscriptaddress, floatfix, showpacs, notitlepage]{revtex4-1}
\pdfoutput=1

\usepackage[utf8]{inputenc}
\usepackage[T1]{fontenc}
\usepackage{lmodern}
\usepackage{graphicx}
\usepackage{amsmath}
\usepackage{amssymb}
\usepackage{wasysym}
\usepackage{bm}
\usepackage{xcolor}
\usepackage[colorlinks, citecolor={blue!50!black}, urlcolor={blue!50!black}]{hyperref}
\usepackage{bookmark}
\usepackage{tabularx}
\usepackage{mathtools}
\usepackage{microtype}
\usepackage[load=physical,load=abbr]{siunitx}

\DeclarePairedDelimiter\abs{\lvert}{\rvert}%
\DeclarePairedDelimiter\norm{\lVert}{\rVert}%
\makeatletter
\let\oldabs\abs
\def\abs{\@ifstar{\oldabs}{\oldabs*}}
\let\oldnorm\norm
\def\norm{\@ifstar{\oldnorm}{\oldnorm*}}
\makeatother

\newcolumntype{L}[1]{>{\raggedright\arraybackslash}p{#1}}
\newcolumntype{C}[1]{>{\centering\arraybackslash}p{#1}}
\newcolumntype{R}[1]{>{\raggedleft\arraybackslash}p{#1}}

\begin{document}

\author{J. Kammhuber}
\author{M.C. Cassidy}
\author{F. Pei}
\affiliation
{QuTech and Kavli Insitute of Nanoscience, Delft University of Technology, 2600 GA Delft, The Netherlands}

\author{M.P. Nowak}
\affiliation
{QuTech and Kavli Insitute of Nanoscience, Delft University of Technology, 2600 GA Delft, The Netherlands}
\affiliation{Current adress: Faculty of Physics and Applied Computer Science, AGH University of Science and Technology, al. A.Mickiewicza 30, 30-059 Krak{\'o}w, Poland}

\author{A. Vuik}
\affiliation
{QuTech and Kavli Insitute of Nanoscience, Delft University of Technology, 2600 GA Delft, The Netherlands}

\author{D. Car}
\affiliation
{QuTech and Kavli Insitute of Nanoscience, Delft University of Technology, 2600 GA Delft, The Netherlands}
\affiliation
{Department of Applied Physics, Eindhoven University of Technology, 5600 MB Eindhoven, The Netherlands}

\author{S.R. Plissard}
\affiliation
{CNRS-Laboratoire d'Analyse et d'Architecture des Systemes (LAAS), Universit\'e de Toulouse, 7 avenue du colonel Roche, F-31400 Toulouse, France}

\author{E.P.A.M. Bakkers}
\affiliation
{QuTech and Kavli Insitute of Nanoscience, Delft University of Technology, 2600 GA Delft, The Netherlands}
\affiliation
{Department of Applied Physics, Eindhoven University of Technology, 5600 MB Eindhoven, The Netherlands}

\author{M. Wimmer}
\affiliation
{QuTech and Kavli Insitute of Nanoscience, Delft University of Technology, 2600 GA Delft, The Netherlands}
\author{L.P. Kouwenhoven}
\email{L.P.Kouwenhoven@tudelft.nl}
\affiliation
{QuTech and Kavli Insitute of Nanoscience, Delft University of Technology, 2600 GA Delft, The Netherlands}

\title{Conductance through a helical state in an InSb nanowire}
\maketitle

\textbf{ The motion of an electron and its spin are generally not coupled. However in a one dimensional (1D) material with strong spin-orbit interaction (SOI) a helical state may emerge at finite magnetic fields,\cite{Streda2003,Pershin2004} where electrons of opposite spin will have opposite momentum. The existence of this helical state has applications for spin filtering and Cooper pair splitter devices\cite{Sato2010,Rashba2016} and is an essential ingredient for realizing topologically protected quantum computing using Majorana zero modes.\cite{alicea2011,Nayak2008,Yuval2010} Here we report electrical conductance measurements of a quantum point contact (QPC) formed in an indium antimonide (InSb) nanowire as a function of magnetic field. At magnetic fields exceeding \SI{3}{T}, the $2e^2/h$ plateau shows a reentrant conductance feature towards $1e^2/h$ which increases linearly in width with magnetic field before enveloping the $1e^2/h$ plateau. Rotating the external magnetic field either parallel or perpendicular to the spin orbit field allows us to clearly attribute this experimental signature to SOI. We compare our observations with a model of a QPC incorporating SOI and extract a spin orbit energy of $\sim\SI{6.5}{meV}$, which is significantly stronger than the SO energy obtained by other methods.}

Spin-orbit interaction is a relativistic effect where a charged particle moving in an electric field $E$ with momentum $k$ and velocity $v =k/m_0$, experiences an effective magnetic field $B_{SO}=(-1/m_0 c)k\times E$ in its rest frame. The magnetic moment of the electron spin, $\mu=eS/m_0$, interacts with this effective magnetic field, resulting in a spin-orbit Hamiltonian $H_{SO} = -\mu.B_{SO}$ that couples the spin to the orbital motion and electric field. In crystalline materials, the electric field arises from a symmetry breaking that is either intrinsic to the underlying crystal lattice in which the carriers move, known as the Dresselhaus SOI,\cite{dresselhaus1955} or an artificially induced asymmetry in the confinement potential due to an applied electric field, or Rashba SOI.\cite{rashba2015} Wurtzite and certain zincblende nanowires possess a finite Dresselhaus SOI, and so the SOI is a combination of both the Rashba and Dresselhaus components. For zincblende nanowires grown along the [111] growth direction the crystal lattice is inversion symmetric, and so only a Rashba component to the spin-orbit interaction is thought to remain.\cite{winkler2003}

Helical states have been shown to emerge in the edge mode of 2D quantum spin hall topological insulators,\cite{Koenig2007,nowack2013} and in quantum wires created in GaAs cleaved edge overgrowth samples.\cite{quay2010} They have also been predicted to exist in carbon nanotubes under a strong applied electric field,\cite{klinovaja2011} RKKY systems,\cite{klinovaja2013topological} and in InAs and InSb semiconducting nanowires where they are essential for the formation of Majorana zero modes. Although the signatures of Majoranas have been observed in nanowire-superconductor hybrid devices,\cite{mourik2012signatures,albrecht2016exponential} explicit demonstration of the helical state in these nanowires has remained elusive. The measurement is expected to show a distinct experimental signature of the helical state - a return to $1e^2/h$ conductance at the $2e^2/h$ plateau in increasing magnetic field as different portions of the band dispersion are probed.\cite{Streda2003,Pershin2004,Rainis2014} While ballistic transport through nanowire QPCs is now standard,\cite{kammhuber2016conductance,heedt2016ballistic} numerical simulations have shown that the visibility of this experimental signature critically depends on the exact combination of geometrical and physical device parameters.\cite{Rainis2014} 

Here we observe a clear signature of transport through a helical state in a QPC formed in an InSb nanowire when the magnetic field has a component perpendicular to the spin-orbit field. We show that the state evolves under rotation of the external magnetic field, disappearing when the magnetic field is aligned with $B_{SO}$. By comparing our data to a theoretical model, we extract a spin orbit energy $E_{SO} = \SI{6.5}{meV}$, significantly stronger than that measured in InSb nanowires by other techniques.

\begin{figure*}[p]
\begin{center}
\includegraphics[width=2.0\columnwidth]{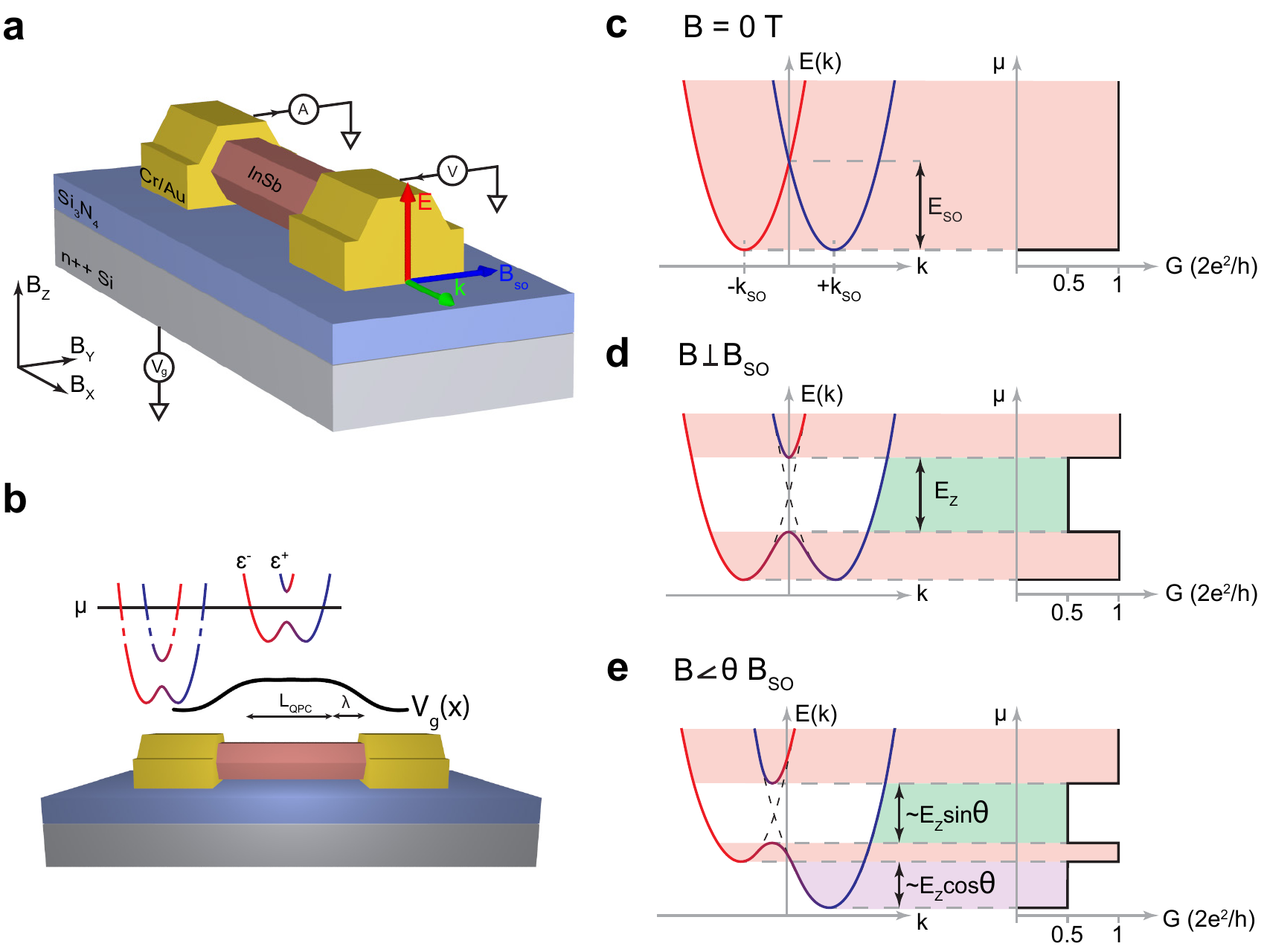}
  \caption{\bf The helical gap in a 1D nanowire device.
  \textbf{a,} \normalfont An InSb nanowire device with a Rashba spin-orbit field $\text{B}_{\text{SO}}$ perpendicular to the wave vector \textbf{k} and the electric field \textbf{E}. A voltage is sourced to one contact, and the resulting conductance measured from the second contact. The chemical potential in the wire, $\mu$, is tuned with a global backgate $\text{V}_\text{g}$. 
  \textbf{b,} \normalfont The QPC channel of length L is defined by the two contacts. The shape of the onset with a lengthscale $\lambda$ is set by the dielectric and screening of the electric field from the metallic contacts resulting in an effective QPC length $L_{QPC}= L- 2\lambda$. 
  \textbf{c,} \normalfont The energy dispersion of the first two subbands for a system with SOI at external magnetic field $B=0$. The SOI causes subbands to shift by $k_{SO}$ in momentum space, as electrons with opposite spins carry opposite momentum.  When the electrochemical potential $\mu$ in the wire is tuned conductance plateaus will occur at integer values of $G_0$.
  \textbf{d,} \normalfont At finite magnetic field $B$ perpendicular to $B_{SO}$, the spin polarized bands hybridize opening a helical gap of size $E_Z$ (green). In this region the conductance reduces from $1\cdot G_0$  to $0.5\cdot G_0$  when $\mu$ is positioned inside the gap.
  \textbf{e,} \normalfont When the magnetic field is orientated at an angle $\theta$ to $B_{SO}$, the size of the helical gap decreases to only include the component of the magnetic field perpendicular to $B_{SO}$. For all angles the reentrant conductance feature at $0.5\cdot G_0$  in the $1\cdot G_0$  plateau will scale linearly with Zeeman energy.}
  \label{Main-fig-1}
\end{center}
\end{figure*}

Figure \ref{Main-fig-1}a shows a schematic image of a typical QPC device. An InSb nanowire is deposited on a degenerately doped silicon wafer covered with a thin (\SI{20}{nm}) SiN dielectric. The QPC is formed in the nanowire channel in a region defined by the source and drain contacts spaced $\sim\SI{325}{nm}$ apart. The chemical potential $\mu$ in the QPC channel, which sets the subband occupation, is controlled by applying a voltage to the gate $V_g$. The electric field in the nanowire, $E$, generated by the backgate and the substrate that the nanowire lies on, both induce a structural inversion asymmetry that results in a finite Rashba spin orbit field. As the wire is translationally invariant along its length, the spin orbit field, $B_{SO}$, is perpendicular to both the electric field and the wire axis. The effective channel length, $L_{QPC}\sim\SI{245}{nm}$, as well as the shape of the onset potential $\lambda\sim\SI{80}{nm}$ are set by electrostatics which are influenced by both the thickness of the dielectric and the amount of electric field screening provided by the metallic contacts to the nanowire (Fig \ref{Main-fig-1}b). Here we report measurements from one device. Data from an additional device that shows the same effect, as well as control devices of different channel lengths and onset potentials, is provided in the Supplementary Information.

The energy-momentum diagrams in Fig \ref{Main-fig-1}c-e show the dispersion from the 1D nanowire model of Refs. 1 and 2 including both SOI with strength $\alpha$ and Zeeman splitting $E_Z=g\mu_B B$, where $g$ is the $g$-factor, $\mu_B$ the Bohr magneton and $B$ the magnetic field strength. These dispersion relations explain how the helical gap can be detected: Without magnetic field, the SOI causes the first two spin degenerate sub-bands to be shifted laterally in momentum space by $\pm k_{SO} = m^* \alpha / \hbar^2$ with $m^*$ the effective electron mass, as electrons with opposite spins carry opposite momentum, as shown in Fig \ref{Main-fig-1}c. The corresponding spin-orbit energy is given by $E_{SO}=\hbar^2 k_{SO}^2/2m^*$. However, here Kramers degeneracy is preserved and hence the plateaus in conductance occur at integer values of $G_0 = 2e^2/h$, as for a system without SOI. Applying a magnetic field perpendicular to BSO the spin bands hybridize and a helical gap, of size $E_Z$ opens as shown in Fig \ref{Main-fig-1}d. When the chemical potential $\mu$ is tuned by the external gate voltage, it first aligns with the bottom of both bands resulting in conductance at $1\cdot G_0$  before reducing from $1\cdot G_0$  to $0.5\cdot G_0$ when $\mu$ is positioned inside the gap. This conductance reduction with a width scaling linearly with increasing Zeeman energy, is a hallmark of transport through a helical state. When the magnetic field is orientated at an angle $\theta$ to $B_{SO}$, the size of the helical gap decreases as it is governed by the component of the magnetic field perpendicular to $B_{SO}$, as shown in Fig \ref{Main-fig-1}e. Additionally, the two sub-band bottoms also experience a spin splitting giving rise to an additional Zeeman gap. For a general angle $\theta$, the QPC conductance thus first rises from 0 to $0.5\cdot G_0$ , then to $1\cdot G_0$ , before dropping to $0.5\cdot G_0$  again. The helical gap thus takes the form of a re-entrant $0.5\cdot G_0$  conductance feature. By comparing to a 1D nanowire model, we can extract both the size of the helical gap $E_{helical}\approx E_Z\sin\theta$ and the Zeeman shift $E_{Zeeman}\approx E_Z\cos\theta$ (see Supplementary Information). This angle dependency is a unique feature of SOI and can be used as a decisive test for its presence in the experimental data.

\begin{figure*}[p]
\begin{center}
\includegraphics[width=2.0\columnwidth]{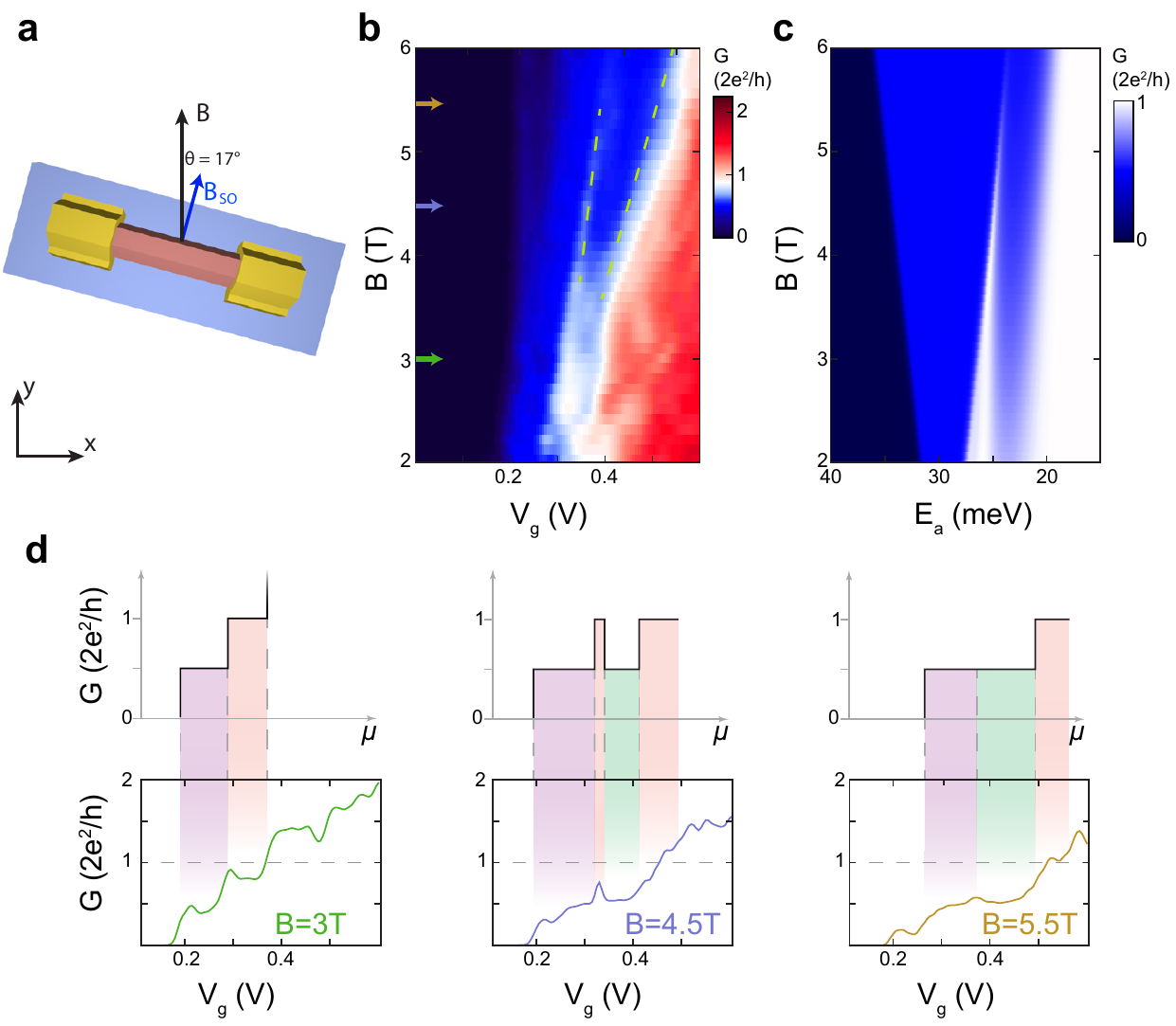}
  \caption{\bf Magnetic field dependence of the helical gap.
  a, \normalfont The nanowire lies in the x-y plane at an angle $\theta =\ang{17}$ relative to the external magnetic field.
  \bf b, \normalfont Differential conductance $dI/dV$ at zero source-drain bias as a function of back gate voltage and external magnetic field. At low magnetic fields conductance plateaus at multiples of $0.5\cdot G_0$  are visible. Above $B=\SI{3}{T}$, a reentrant conductance feature at $0.5\cdot G_0$  appears in the $1\cdot G_0$  plateau. The feature evolves linearly with Zeeman energy indicated by dashed green lines.
  \bf c, \normalfont Numerical simulations of the differential conductance as a function of the potential $\text{E}_{\text{a}}$ and external magnetic field for $L=\SI{325}{nm}$, $\theta = \ang{17}$ and $l_{SO}=\SI{20}{nm}$ (See Supplementary Information for a more detailed description of the model). In the numerical simulations, the conductance plateaus have a different slope compared to the experimental data as the calculations neglect screening by charges in the wire. 
  \bf d, \normalfont Line traces of the conductance map in \textbf{b}. As the helical gap is independent of disorder or interference effects, these and other anomalous conductance features average out in a 2D colorplot improving the visibility of the helical gap in b compared to the individual traces in \textbf d.
  }
  \label{Main-fig-2}
\end{center}
\end{figure*}

\begin{figure*}[p]
\begin{center}
\includegraphics[width=2.0\columnwidth]{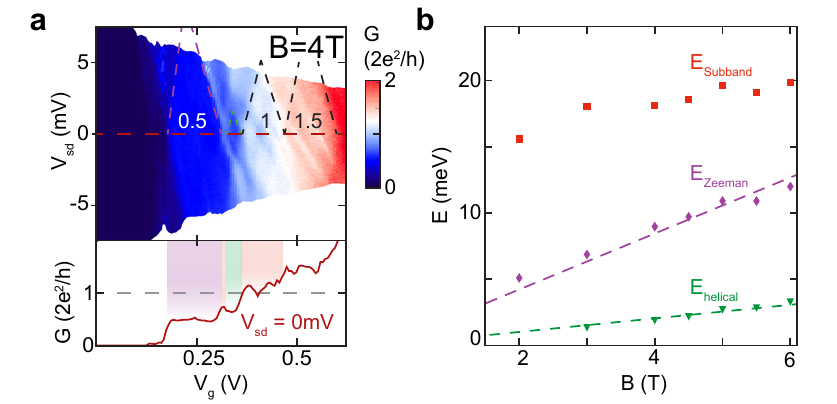}
  \caption{\bf Voltage bias spectroscopy of the helical gap.
  a, \normalfont Conductance measurement as a function of QPC gate and source-drain bias voltage at $B=\SI{4}{T}$. The observed helical gap (green) is a stable feature in voltage bias. Dotted lines are drawn as guide to the eye indicating the plateau edges.
  \bf b. \normalfont Evolution of the energy levels extracted from scans similar to \textbf{a, } at increasing magnetic field. Fits with intercept fixed at zero (dotted lines) give the $g$-factor of the first subband and the offset angle via $g=1/(\mu_B\cos\theta)\cdot dE/dB$ and $E_{helical}/E_{Zeeman}\approx\tan\theta$. We find $g = 38 \pm 1$ and $\theta = \ang{13}\pm\ang{2}$. Individual scans are included in the Supplementary Information.}
  \label{Main-fig-3}
\end{center}
\end{figure*}

Figure \ref{Main-fig-2} shows the differential conductance $dI/dV$ of our device at zero source-drain bias as a function of gate and magnetic field. Here the magnetic field $B$ is offset at a small angle $\theta = \ang{17}$ from $B_{SO}$ (see Fig \ref{Main-fig-2}a). We determine that our device has this orientation from the angle-dependence of the magnetic field, by clearly resolving the $1\cdot G_0$  plateau before the re-entrant conductance feature, which is reduced at larger angles (see Supplementary Information).  For low magnetic fields, we observe conductance plateaus quantized in steps of $0.5\cdot G_0$ , as typical for a QPC in a spin polarizing B-field with or without SOI. Above $B = \SI{3}{T}$, the $1\cdot G_0$  plateau shows a conductance dip to $0.5\cdot G_0$ . This reentrant conductance feature evolves continuously as a function of magnetic field, before fully enveloping the $1\cdot G_0$  plateau for magnetic fields larger than around \SI{5.5}{T}. Line traces corresponding to the colored arrows in Fig \ref{Main-fig-2}b are shown in Fig \ref{Main-fig-2}d. The feature is robust at higher temperatures up to 1K, as well across multiple thermal cycles (see Supplementary Information). 
Using the 1D nanowire model with $\theta = \ang{17}$ we find that the helical gap feature vanishes into a continuous $0.5\cdot G_0$ plateau when $E_Z>2.4 E_{SO}$. Using the extracted $g$-factor $g=38$ of our device (see Fig \ref{Main-fig-3} and Supplementary Information) we find a lower bound for the spin-orbit energy $E_{SO}=\SI{5.5}{meV}$, corresponding to a spin-orbit length $l_{SO}=1/k_{SO}\approx\SI{22}{nm}$. For a second device, we extract a similar value $E_{SO}=\SI{5.2}{meV}$. Recently it has been highlighted that the visibility of the helical gap feature depends crucially on the shape of the QPC potential.18 To verify that our observation is compatible with SOI in this respect, we perform self-consistent simulations of the Poisson equation in Thomas-Fermi approximation for our device geometry. The resulting electrostatic potential is then mapped to an effective 1D QPC potential for a quantum transport simulation using parameters for InSb (for details, see Supplementary Information). These numerical simulations, shown in Fig \ref{Main-fig-2}c, fit best for $l_{SO}=\SI{20}{nm}$ ($E_{SO} = \SI{6.5}{meV}$) and agree well with the experimental observation, corroborating our interpretation of the re-entrant conductance feature as the helical gap.

Voltage bias spectroscopy, as shown in Fig \ref{Main-fig-3}a confirms that this state evolves as a constant energy feature. By analyzing the voltage bias spectroscopy data at a range of magnetic fields, we quantify the development of the initial $0.5\cdot G_0$  plateau, as well as the reentrant conductance feature (Fig \ref{Main-fig-3}b). From the evolution of the width of the first $0.5\cdot G_0$  plateau, we can calculate the $g$-factor of the first sub-band $g=38\pm1$. This number is consistent with the recent experiments, which reported g factors of $35-50$.\cite{van2012quantized,nadj2012spectroscopy} Comparing the slopes of the Zeeman gap and the helical gap $E_h/E_Z\approx\tan\theta$ provides an alternative way to determine the offset angle $\theta$. We find $\theta=\ang{13}\pm\ang{2}$ which is in reasonable agreement with the angle determined by magnetic field rotation.

\begin{figure*}[p]
\begin{center}
\includegraphics[width=2.0\columnwidth]{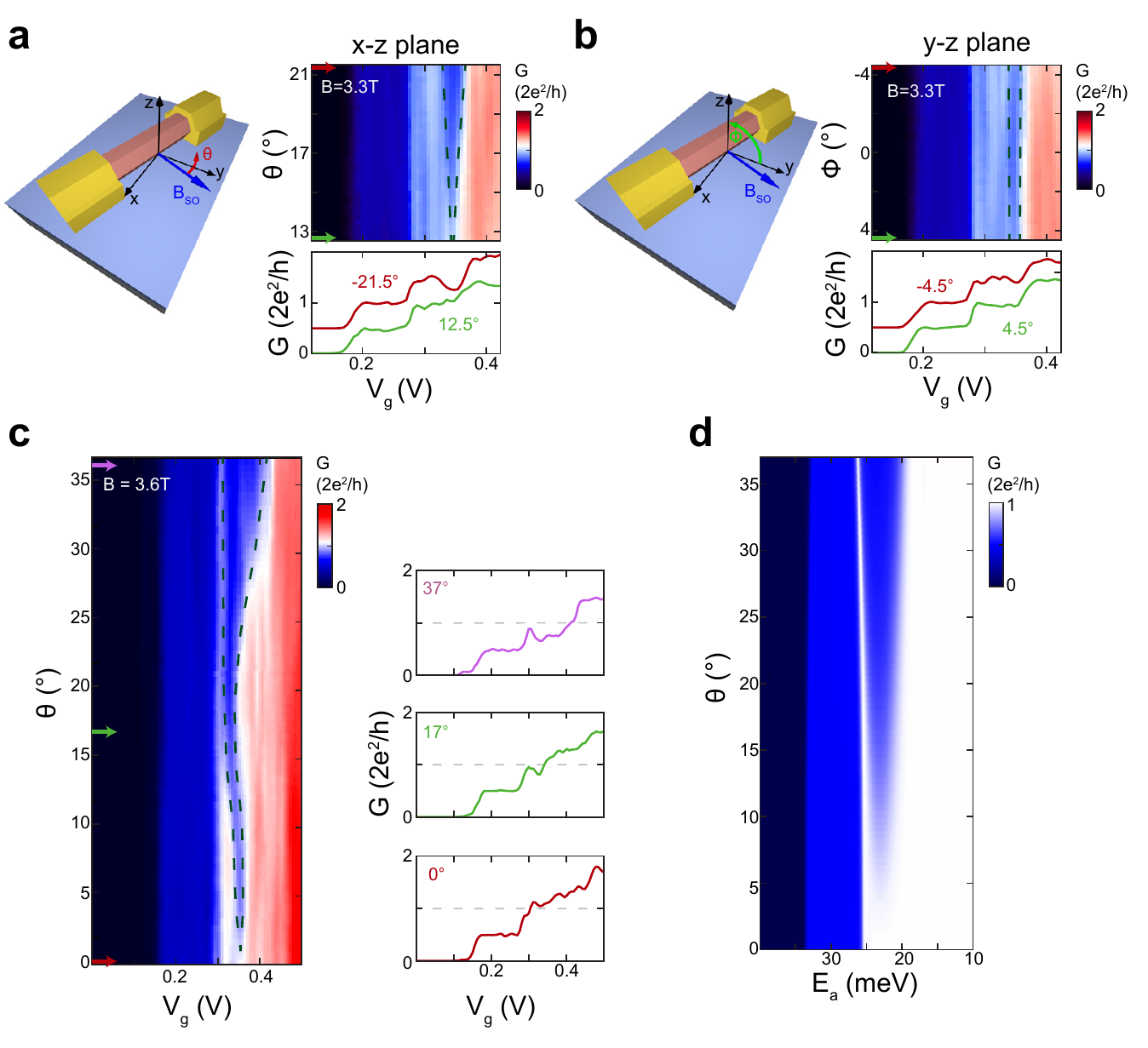}
  \caption{\bf Angle dependence of the helical gap.
  a, \normalfont Rotation of the magnetic field at $B=\SI{3.3}{T}$ in the x-y plane parallel to the substrate shows strong angle dependence of the helical gap. The conductance dip closes when $B$ is rotated towards $B_{SO}$ and opens when $B$ is rotated away from $B_{SO}$.
  \bf b, \normalfont Rotation of the magnetic field at $B=\SI{3.3}{T}$ in the y-z plane, mostly perpendicular to $B_{SO}$. While the angle range is identical to \textbf{a} there is little change in the conductance dip. 
  \bf c, \normalfont Rotation of the magnetic field at $B=\SI{3.6}{T}$ in the x-y plane over a large angle range. The conductance dip disappears when $B$ is parallel to $B_{SO}$ which gives the exact offset angle between $B_{SO}$ and $B_Z$, $\theta=\ang{17}$.
  \bf d, \normalfont Numerical simulations of the differential conductance in a magnetic field rotated along $\theta$ in the x-y plane with $L=\SI{325}{nm}$ and $l_{SO}=\SI{20}{nm}$.}
  \label{Main-fig-4}
\end{center}
\end{figure*}

To confirm that the reentrant conductance feature agrees with spin orbit theory, we rotate the magnetic field in the plane of the substrate at a constant magnitude $B = \SI{3.3}{T}$, as shown in Fig \ref{Main-fig-4}a. When the field is rotated towards being parallel to $B_SO$, the conductance dip closes, while when it is rotated away from $B_{SO}$, the dip increases in width and depth. In contrast, when the magnetic field is rotated the same amount around the y-z plane, which is largely perpendicular to $B_{SO}$, there is little change in the reentrant conductance feature, as shown in Fig \ref{Main-fig-4}b. Figure \ref{Main-fig-4}c shows the result of rotating through a larger angle in the x-y plane shows this feature clearly evolves with what is expected for spin orbit. Our numerical simulations in Figure \ref{Main-fig-4}d agree well with the observed experimental data. The small difference in the angle evolution between the numerical simulations and experimental data can be attributed to imperfect alignment of the substrate with the x-y plane. 

The extracted SO energy of \SI{6.5}{meV} is significantly larger than that obtained via other techniques, such as weak anti localization (WAL) measurements,\cite{van2015spin} and quantum dot spectroscopy.\cite{nadj2012spectroscopy} This is not entirely unexpected, due to the differing geometry for this device and different conductance regime it is operated in. Quantum dot measurements require strong confinement, and so the Rashba SOI is modified by the local electrostatic gates used to define the quantum dot. Weak anti-localization measurements are performed in an open conductance regime, however they assume transport through a diffusive, rather than a ballistic channel. Neither of these measurements explicitly probe the spin orbit interaction where exactly one mode is transmitting in the nanowire, the ideal regime for Majoranas, and so the spin orbit parameters extracted from QPC measurements offer a more accurate measurement of the SOI experienced by the Majorana zero mode. Also, the SOI in a nanowire can be different for every subband, and it is expected that the lowest mode has the strongest spin-orbit due to a smaller confinement energy.\cite{winkler2003} Additionally, the finite diameter of the nanowire, together with impurities within the InSb crystal lattice both break the internal symmetry of the crystal lattice and may contribute a non-zero Dresselhaus component to the spin orbit energy that has not been previously considered.
While high quality quantized conductance measurements have been previously achieved in short channel devices\cite{kammhuber2016conductance} ($L\sim\SI{150}{nm}$), the channel lengths required for observing the helical gap are at the experimental limit of observable conductance quantization. As shown in the Supplementary Information, small changes in the QPC channel length, spin-orbit strength or the QPC potential profile are enough to obscure the helical gap, particularly for wires with weaker SOI. We have fabricated and measured a range of QPCs with different length and potential profiles, and only two devices of $L\sim\SI{300}{nm}$ showed unambiguous signatures of a helical gap. 

Several phenomena have been reported to result in anomalous conductance features in a device such as this. Oscillations in conductance due to Fabry-Perot resonances are a common feature in clean QPCs. Typically the first oscillation at the front of each plateau is the strongest and the oscillations monotonically decrease in strength further along each plateau.\cite{Rainis2014,van2015spin}  In our second device, we clearly observe Fabry-Perot conductance oscillations at the beginning of each plateau, however these oscillations are significantly weaker than the subsequent conductance dip. Furthermore we observe Fabry-Perot oscillations at each conductance plateau, while the reentrant conductance feature is only present at the $1\cdot G_0$  plateau. Additionally, the width of the Fabry-Perot oscillations does not change with increasing magnetic field, unlike the observed re-entrant conductance feature. A local quantum dot in the Coulomb or Kondo regimes can lead to conductance suppression, which increases in magnetic field.\cite{heyder2015relation} However both effects should be stronger in the lower conductance region, and exists at zero magnetic field, unlike the feature in our data. Additionally, a Kondo resonance should scale with $V_{sd}=\pm g\mu_B B/e$ as a function of external magnetic field, decreasing instead of increasing the width of the region of suppressed conductance. Given the g factor measured in InSb quantum dots, and its variation with the angle of applied magnetic field $g=35-50$,\cite{nadj2012spectroscopy} we can exclude both these effects. Similarly the Fano effect and disorder can also induce a conductance dip, but these effects should not increase linearly with magnetic field. The 0.7 anomaly occurs at the beginning of the plateau, and numerical studies have shown it does not drastically affect the observation of the helical gap.\cite{Goulko2014} In conclusion, we have observed a return to $1e^2/h$ conductance at the $2e^2/h$ plateau in a QPC in an InSb nanowire. The continuous evolution in increasing magnetic field and the strong angle dependence in magnetic field rotations agree with a SOI related origin of this feature and distinguish it from Fabry-Perot oscillations and other g-factor related phenomena. Additional confirmation is given by numerical simulations of an emerging helical gap in InSb nanowires. The extracted spin orbit energy of \SI{6.5}{meV} is significantly larger than what has been found by other techniques, and more accurately represents the true spin orbit energy in the first conduction mode. Such a large spin orbit energy reduces the requirements on nanowire disorder for reaching the topological regime,\cite{sau2012experimental} and offers promise for using InSb nanowires for the creation of topologically protected quantum computing devices.

\section*{Methods}

\subsection*{Device Fabrication}
The InSb nanowires were grown using the metalorganic vapor phase epitaxy (MOVPE) process.28 The InSb nanowires were deposited using a deterministic deposition method on a degenerately doped silicon wafer. The wafer covered with \SI{20}{nm} of low stress LPCVD SiN which is used as a high quality dielectric. Electrical contacts (Cr/Au, \SI{10}{nm}/\SI{110}{nm}) defined using ebeam lithography were then evaporated at the ends of the wire. Before evaporation the wire was exposed to an ammonium polysulfide surface treatment and short helium ion etch to remove the surface oxide and to dope the nanowire underneath the contacts.\cite{kammhuber2016conductance}

\subsection*{Measurements}
Measurements are performed in a dilution refrigerator with base temperature $\sim\SI{20}{mK}$ fitted with a 3-axis vector magnet, which allowed for the external magnetic field to be rotated in-situ. The sample is mounted with the substrate in the x-y plane with the wire orientated at a small offset angle $\theta = \ang{17}$ from the x-axis. We measure the differential conductance $G=dI/dV$ using standard lock-in techniques with an excitation voltage of \SI{60}{\micro\volt} and frequency $f=\SI{83}{Hz}$. Additional resistances due to filtering are subtracted to give the true conductance through the device.

\subsection*{Numerical transport simulations}
We use the method of finite differences to discretize the one-dimensional nanowire model of Ref \citenum{Pershin2004}. In order to obtain a one-dimensional QPC potential, we solve the Poisson equation self-consistently for the full three-dimensional device structure treating the charge density in the nanowire in Thomas-Fermi approximation. To this end, we use a finite element method, using the software FEniCS.\cite{logg2012automated} The resulting three-dimensional potential is then projected onto the lowest nanowire subband and interpolated using the QPC potential model of Ref \cite{Rainis2014}.  Transport in the resulting tight-binding model is calculated using the software Kwant.\cite{groth2014kwant}

\begin{acknowledgements}
We gratefully acknowledge D. Xu, \"O. G\"ul, S. Goswami, D. van Woerkom and R.N. Schouten for their technical assistance and helpful discussions. This work has been supported by funding from the Netherlands Foundation for Fundamental Research on Matter (FOM), the Netherlands Organization for Scientific Research (NWO/OCW), the Office of Naval Research, Microsoft Corporation Station Q, the European Research Council (ERC) and an EU Marie-Curie ITN. 
\end{acknowledgements}

\section*{Author contributions}
J.K and F.P. fabricated the samples, J.K. M.C.C. and F.P. performed the measurements with input from M.W.. M.W., M.N. and A.V. developed the theoretical model and performed simulations. D.C., S.R.P. and E.P.A.M.B. grew the InSb nanowires. All authors discussed the data and contributed to the manuscript.

\newpage

\clearpage
\setcounter{figure}{0}
\setcounter{page}{1}
\onecolumngrid
\renewcommand{\theequation}{S\arabic{equation}}
\renewcommand{\thefigure}{S\arabic{figure}}
\renewcommand{\bibnumfmt}[1]{[S#1]}
\renewcommand{\citenumfont}[1]{S#1}

\begin{center}
\textbf{\large Supplementary Information for: Conductance through a helical state in an InSb nanowire}
\end{center}

\begin{center}
\normalsize{
J.~Kammhuber,$^{1}$ M.~C.~Cassidy,$^{1}$ F.~Pei,$^{1}$ M.~P.~Nowak,$^{1,2}$ A.~Vuik,$^{1}$ D.~Car,$^{1,3}$

S.~R.~Plissard,$^{4}$ E.~P.~A.~M.~Bakkers,$^{1,3}$ M.~Wimmer,$^{1}$ Leo~P.~Kouwenhoven$^{1,*}$}
\end{center}
\small
\begin{center}
\emph{$^\mathit{1}$QuTech and Kavli Institute of Nanoscience, Delft University of Technology, 2600 GA Delft, The Netherlands}

\emph{$^\mathit{2}$ Current adress: Faculty of Physics and Applied Computer Science, AGH University of Science and Technology, al. A.Mickiewicza 30, 30-059 Krak{\'o}w, Poland}

\emph{$^\mathit{3}$Department of Applied Physics, Eindhoven University of Technology, 5600 MB Eindhoven, The Netherlands}

\emph{$^\mathit{4}$CNRS-Laboratoire d'Analyse et d'Architecture des Systemes (LAAS), Universit\'e de Toulouse, 7 avenue du colonel Roche, F-31400 Toulouse, France}
\end{center}
\large

\section{Numerical simulations of the conductance through a helical states}

\subsection{Poisson calculations in a 3D nanowire device}

Observing the helical gap in a semiconducting nanowire crucially depends on the smoothness of the electrostatic potential profile between the two contacts \cite{supp1}. When the potential profile changes too abruptly, it forms a tunnel barrier which suppresses conductance well below quantized values, thereby masking features of the helical gap. On the other hand, if the potential varies on a length scale much larger than the characteristic spin-orbit coupling length $l_{SO}$, transmission through the `internal state' (the smaller-momentum state of the two right-moving states in the bottom of the lower band) is suppressed. This reduces the first $2e^2/h$ plateau in the conductance to a $1e^2/h$ plateau, thereby concealing again the helical gap.

Because of the crucial role of the electrostatic potential, we perform realistic Poisson calculations to compute the potential $\phi(\vec{r})$ in the nanowire (with $\vec{r} = (x, y, z)$), solving the Poisson equation of the general form
\begin{equation}
\nabla^2 \phi(\vec{r}) = -\frac{\rho(\vec{r})}{\epsilon},
\label{eq:poisson}
\end{equation}
with $\epsilon$ the dielectric permittivity and $\rho$ the charge density. For the charge density $\rho$, we apply the Thomas-Fermi approximation~\cite{supp2}
\begin{equation}
\rho(\vec{r}) = \frac{e}{3\pi^2 \epsilon} \left(\frac{2m^* e \phi(\vec{r})}{\hbar^2}\right)^{3/2},
\label{eq:ThomasFermi}
\end{equation}
where $m^*$ is the effective mass of InSb. \\
For a given charge density $\rho$, we solve Eq.~\ref{eq:poisson} numerically for the potential using the finite element package FEniCS \cite{supp3}. We model the two normal contacts as metals with a fixed potential $V_\text{N} = 0.22$~V, assuming a small work function difference between the nanowire and the normal contacts. The back gate is modeled as a fixed potential $V_\text{G}$ along the bottom surface of the dielectric layer. We use the dielectric permittivities for InSb and SiN in the wire and the dielectric layer respectively. The FEM mesh, with its dimensions and boundary conditions, is depicted in Fig.~\ref{fig_supp_t1}a. \\
We apply the Anderson mixing scheme~\cite{supp4} to solve the nonlinear equation formed by Eqs.~\ref{eq:poisson} and~\ref{eq:ThomasFermi} self-consistently. An example of a self-consistent Poisson potential with Thomas-Fermi density is plotted in Fig.~\ref{fig_supp_t1}b.
\subsection{Conductance calculations in a 1D model with a projected potential barrier}

\begin{figure}
\includegraphics[width=0.9\columnwidth]{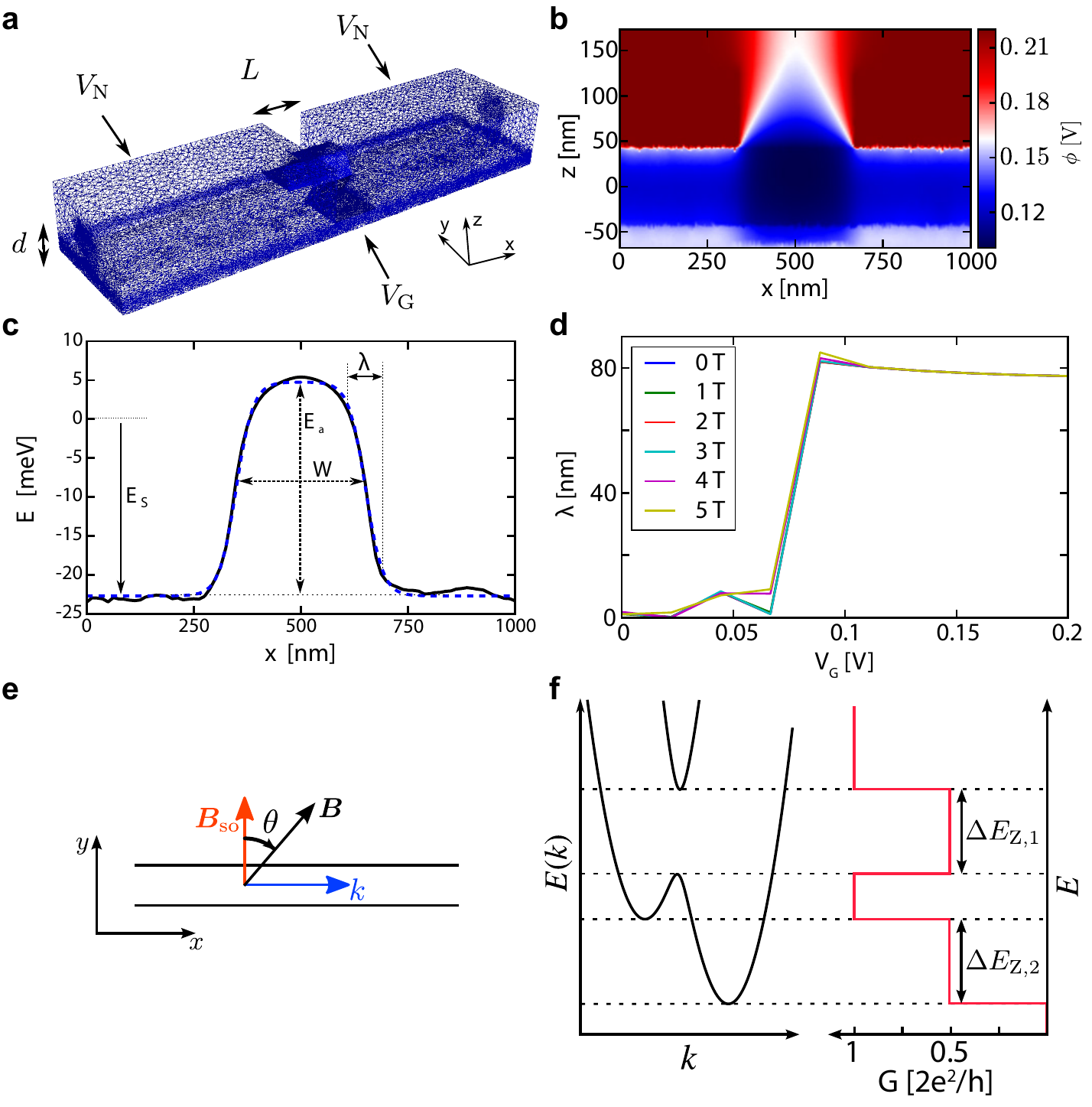}
\caption{\bf{a, }\normalfont Example of a finite element mesh used for 3D Poisson calculations. $L$ denotes the QPC length (spacing between the two contacts), $d$ the thickness of the dielectric layer, which is set to \SI{20}{nm}. $L$ is set to \SI{325}{nm} for the simulations in the main text, and varied from 175 to \SI{425}{nm} to show the length dependence of the helical gap feature in Fig \ref{fig_supp_e6}. The two boundary conditions applied are a potential $V_\text{N}$ on the contacts and a potential $V_\text{G}$ underneath the dielectric layer. The mesh between the two contacts is left out for visibility purposes.
\bf{b, }\normalfont Cross section plot of the 3D Poisson potential for $V_\text{G} = \SI{0.156}{V}$ and $V_\text{N} = \SI{0.22}{V}$. The cross section is taken along the wire axis (x-axis) for fixed $y = \SI{0}{nm}$ in the middle of the wire. The effective QPC length runs from $\sim 340$ to $\sim \SI{660}{nm}$. The nanowire is situated between $z = \SI{-50}{nm}$ and $z=\SI{50}{nm}$.
\bf{c, }\normalfont Projected potential $\hat{\phi}(x)$ (black curve) and fitted potential $V(x)$ (blue dashed curve) for $V_\text{G} = \SI{0.156}{V}$, corresponding to the potential of Fig.~\ref{fig_supp_t1}b. Indicated are the fitting parameters $E_\text{s}$, $E_\text{a}$, $W$ and $\lambda$ of the function Eq. \ref{eq:tanh_fit}.
\bf{d, }\normalfont The fitting parameter $\lambda$ as a function of back gate voltage $V_\text{G}$. Different colors denote different magnetic field strengths $B$. A jump in $\lambda \approx 0$ (abrupt step potential) to $\lambda \approx \SI{80}{nm}$ occurs when charge enters the wire, screening the electric field.
\bf{e, }\normalfont Rashba nanowire in an external magnetic field: the one-dimensional nanowire is oriented along the x-axis, and the spin-orbit field $\mathbf{B}_\text{so}$ perpendicular, along the $y$-axis. The external magnetic field $\mathbf{B}$ forms an angle $\theta$ with respect to $\mathbf{B}_\text{so}$. 
\bf{f, }\normalfont Schematic picture of the band structure $E(k)$ of a Rashba nanowire in a magnetic field (left panel) and the corresponding quantum point conductance $G$ (right panel).
}
\label{fig_supp_t1}
\end{figure}

To apply the 3D Poisson potential in a simple 1D nanowire model, we convert 
the three-dimensional potential $\phi(x, y, z)$ to a one-dimensional effective potential barrier $\hat{\phi}(x)$ by projecting $\phi$ on the transverse wave functions $\psi(y, z)$ in the nanowire:
\begin{equation}
\hat{\phi}(x) = \langle \psi(y, z) | \phi(x, y, z) | \psi(y, z) \rangle.
\label{eq:potprojection}
\end{equation}
To do this, we compute the eigenenergies of the Hamiltonian of a two-dimensional cross section at a point $x_0$ along the wire, with a corresponding potential $\phi(x_0, y, z)$. The effective potential barrier is then given by the ground state of the Hamiltonian. 
The longitudinal variation of the potential barrier is obtained by computing the ground state of the transverse Hamiltonian at many points along the wire. An example of the projected potential is given in Fig.~\ref{fig_supp_t1}c with the solid-black curve.

Due to rough boundary conditions in the FEM mesh (see the edges of the dielectric layer and the normal contacts in the potential of Fig.~\ref{fig_supp_t1}b), the projected potential $\hat{\phi}(x)$ shows some roughness that may cause unwanted scattering events (see black curve in Fig.~\ref{fig_supp_t1}c). To avoid this, we fit $\hat{\phi}(x)$ to a linear combination of hyperbolic tangents, given by
\begin{equation}
V(x) = \frac{E_\mathrm{a}}{2}\left[ \mathrm{tanh}\left(\frac{x - x_\text{s} + W/2}{\lambda/2} \right) \right.
- \left. \mathrm{tanh}\left(\frac{x - x_\text{s} - W/2}{\lambda/2} \right) \right] 
+ E_\text{s}.
\label{eq:tanh_fit}
\end{equation}
Here, $E_\text{a}$ is the amplitude, $W$ the width and $E_\text{s}$ the downshift in energy of the potential barrier, which varies along $x$ on a typical length scale $\lambda$, as indicated in Fig.~\ref{fig_supp_t1}c. The horizontal shift of the barrier to the middle of the nanowire is denoted by $x_\text{s}=500$~nm.

The parameter $\lambda$ expresses the smoothness of the barrier. We find that $\lambda$ is close to zero when no charge is present in the wire and the boundary conditions result in an abrupt step in the potential between the contacts and the uncovered part of the wire. When charge enters the wire, it screens the electric field, thereby smoothening the potential. 
For a QPC length of 325~nm we find in this regime $\lambda \approx 80$ nm. The value of $\lambda$ is reduced for smaller QPC lengths, but saturates to $\lambda \approx 80$ nm for longer QPC lengths. Moreover we find that $\lambda$ has only a little dependency on the back gate voltage $V_\text{G}$ or the applied magnetic field $B$ (Fig.~\ref{fig_supp_t1}d). Taking advantage of the latter and the fact that we are interested in the conductance of the wire in the vicinity of the helical-gap feature -- where the screening {\it is} present -- we assume $\lambda$ constant in $V_\text{G}, B$ space for the conductance calculation.

For the conductance calculations we consider transport through a two-mode nanowire described by the Hamiltonian
\begin{equation}
\mathcal{H} = \left[ \frac{\hbar^2 k_x^2}{2m^*} + V(x) \right] \sigma_0 + \alpha \sigma_y k_x + \\
\frac{1}{2} g \mu_B B (\sigma_x \sin \theta + \sigma_y \cos \theta),
\label{eq:1DHamiltonian}
\end{equation}
where $\sigma$ denote the Pauli matrices (with $\sigma_0$ the identity matrix) and $V(x)$ is fit to the projected potential barrier, as expressed in Eqs.~\ref{eq:potprojection} and \ref{eq:tanh_fit}. Spin-orbit coupling strength is given by $\alpha = \hbar^2/m^*l_{SO}$ where $l_{SO}$ we use as a free parameter. We take the effective mass $m^* = 0.014 m_0$ of InSb and $g = -38$ (unless stated otherwise) as estimated in the main text. Note that for the coordinate system used here, where the wire lies along the x direction and $\theta$ is the angle between $B_{SO}$ and the external magnetic field. The Hamiltonian Eq.~\ref{eq:1DHamiltonian} is discretized on a mesh with lattice spacing $\Delta x = 4$ nm. Assuming translational invariance of the boundary conditions at the ends of the wire one arrives at the scattering problem that is solved using the Kwant package~\cite{supp5} to obtain the linear-response conductance within the Landauer-B{\"u}ttiker formalism.

\section{Angle-dependence of conductance in Rashba nanowires}

\subsection{Theoretical model}

We consider a one-dimensional nanowire with Rashba spin-orbit
interaction (SOI) in an external magnetic field $\mathbf{B}$. The
field $\mathbf{B}$ is oriented at an angle $\theta$ with respect to
the effective magnetic field $\mathbf{B}_\text{so}$ due to Rashba SOI,
as shown in Fig.~\ref{fig_supp_t1}e. This setup is described by the
Hamiltonian:\cite{supp6}
\begin{equation}\label{eq:Hamiltonian}
H=\frac{p^2}{2m^*} + \frac{\alpha}{\hbar} p \sigma_y + \frac{1}{2} E_\text{Z}
\left(\sin(\theta)\sigma_x + \cos(\theta) \sigma_y\right)\,.
\end{equation}
In this expression, $p$ is the momentum operator, $m^*$ is the effective mass,
$\alpha$ the Rashba SOI-strength, and $\sigma_{x,y}$ the Pauli matrices.
The Zeeman energy $E_Z=g \mu_B B$, where $g$ is the g-factor,
and $\mu_\text{B}$ the Bohr magneton. In Eq.~\eqref{eq:Hamiltonian} we
assumed without loss of generality a magnetic field in the x-y-plane; the
band structure however only depends on the relative angle $\theta$ of
$\mathbf{B}$ with $\mathbf{B}_\text{so}$.

The Rashba SO-strength $\alpha$ can be related to an effective length scale,
the spin-orbit length
\begin{equation}
l_\text{so}=\frac{\hbar^2}{m \alpha}
\end{equation}
and to an energy scale, the spin-orbit energy
\begin{equation}
E_\text{so}=\frac{m \alpha^2}{2 \hbar^2}\,.
\end{equation}

Defining length in units of $l_\text{so}$ and energy in units of $E_\text{so}$
it is possible to write the Hamiltonian in a convenient dimensionless
form:
\begin{equation}
H=\frac{d^2}{dx^2} + 2 \frac{d}{dx} \sigma_y + \frac{1}{2} \frac{E_Z}{E_\text{so}} (\sin(\theta) \sigma_x
+\cos(\theta) \sigma_y)\,.
\end{equation}
Proper units will be restored in the final result.

In an translationally invariant nanowire, the wave vector $k$ is a good
quantum number and the Rashba Hamiltonian is readily diagonalized as
\cite{supp6}
\begin{align}
E_\pm(k) = k^2 \pm \frac{1}{2} \sqrt{\left(\frac{E_\text{Z}}{E_\text{so}}\right)^2 + 16 k^2 + 8 \frac{E_\text{Z}}{E_\text{so}} k \cos(\theta)}\,.
\end{align}

The resulting band structure for a general angle $\theta$ is shown
schematically in the left panel of Fig.~\ref{fig_supp_t1}f. The band
structure can be related to an idealized quantum point contact (QPC) conductance
by counting the number of propagating modes at a given energy $E$ (see
right panel of Fig.~\ref{fig_supp_t1}f).

In the following we will derive from the band structure: (i) the size of the
$1 e^2/h$ plateaus in energy (denoted by $\Delta E_{Z,1}$ and
$\Delta E_{Z,2}$). This is directly measurable using the finite bias
dependence of the QPC conductance (measuring so-called QPC diamonds). (ii)
The critical field for which the spin-orbit induced $2 e^2/h$ conductance
(the size of this plateau is denoted as $\Delta E_\text{so}$) vanishes. This
allows for an estimate of the spin-orbit strength from the magnetic field
dependence in experiment.

\subsection{Size of Zeeman-induced gaps}

In order to compute the size of the different QPC plateaus in
energies, we need to compute the value of the minima and maxima of the
bands $E_\pm(k)$.  This can be done exactly using a computer algebra
program (we used Mathematica), as it only involves solving for the
roots of polynomials up to fourth order. The resulting expressions are
however quite cumbersome, and it is more useful to find an approximate
expression doing a Taylor approximation.  Up to second order in
$E_z/E_\text{so}$ we then find the simple expressions
\begin{align}
\Delta E_{\text{Z},1} & \approx E_Z \sin\theta\,,\\
\Delta E_{\text{Z},2} & \approx E_Z \cos\theta\,.
\end{align}

\subsection{Critical magnetic field for the spin-orbit induced $2 e^2/h$-plateau}

The spin-orbit induced $2 e^2/h$ region persists only up to a critical
Zeeman splitting $E_\text{Z,crit}$, after which the two $1e^2/h$-plateaus
merge into one. In the band structure, this corresponds to a transition
from three extrema in $E_-(k)$ (two minima, one maximum) to only one
minimum. The critical Zeeman splitting where this happens can be solved for
exactly using Mathematica:

\begin{equation}
\frac{E_\text{Z,crit}}{E_\text{so}} = \sqrt{\frac{54 \cos (8 \theta
    )+3 M_1^{\frac{2}{3}}+6 \left(3 M_1^{\frac{1}{3}}-4\right) \cos (4
    \theta )-2 M_1^{\frac{1}{3}}-30}{M_2^\frac{1}{3}}}
\end{equation}
where
\begin{align}
  M_1 & = 68 - 86 \cos (4 \theta )-36 \cos (8 \theta )+54 \cos (12 \theta )+512 \sqrt{\sin ^4(2 \theta ) \cos ^2(2 \theta )}\\
  M_2 & = 68 -86 \cos (4 \theta )-36 \cos (8 \theta )+54 \cos (12 \theta )+ 256 \sqrt{\sin ^2(2 \theta ) \sin ^2(4 \theta )}
\end{align}

For $\theta=17^\circ$ this gives $E_\text{Z,crit}=2.386 E_\text{so}$
and for $\theta=10^\circ$ $E_\text{Z,crit} = 2.695 E_\text{so}$. When
the value of the nanowire g-factor is extracted from experiment, the critical
Zeeman splitting can be translated into a critical magnetic field. The
magnetic field up to which the spin-orbit induced $2e^2/h$-plateau is still
visible in experiment can then be used to set a \emph{lower bound} on the
spin-orbit energy. It is a lower bound, as for a given QPC potential the
$2e^2/h$ may not be visible any more despite in principle being
present in the band structure. A more detailed transport calculation can
be used to improve on this bound.

\newpage
\section{Device 1 - Additional Data}

\begin{figure*}[!h]
\begin{center}
\includegraphics[width=0.75\columnwidth]{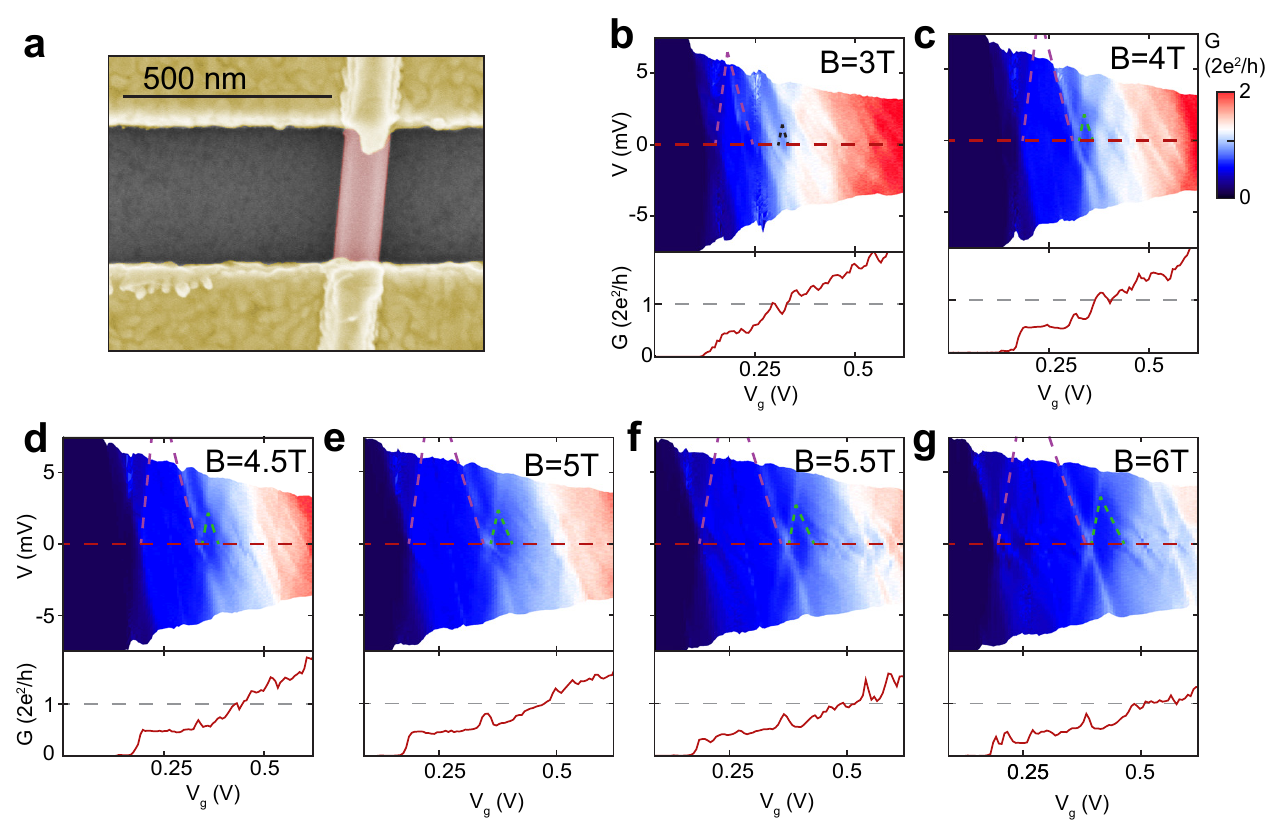}
\caption{\bf{Voltage bias spectroscopy. }\normalfont
\bf{a, }\normalfont False color SEM image of device 1. The InSb nanowire is shown in red and Cr/Au contacts in yellow.
\bf{b-g, }\normalfont Conductance measurements as a function of QPC gate voltage $\text{V}_g$ and source-drain bias voltage $\text{V}_{sd}$ at increasing magnetic field. Dotted lines indicate the helical gap as well as the $0.5\cdot\text{G}_0$ plateau. The helical gap shows as feature stable in $\text{V}_{sd}$ and evolves linearly with magnetic field.
}
\label{fig_supp_e1}
\end{center}
\end{figure*}

\begin{figure}[!h]
\includegraphics[width=0.75\columnwidth]{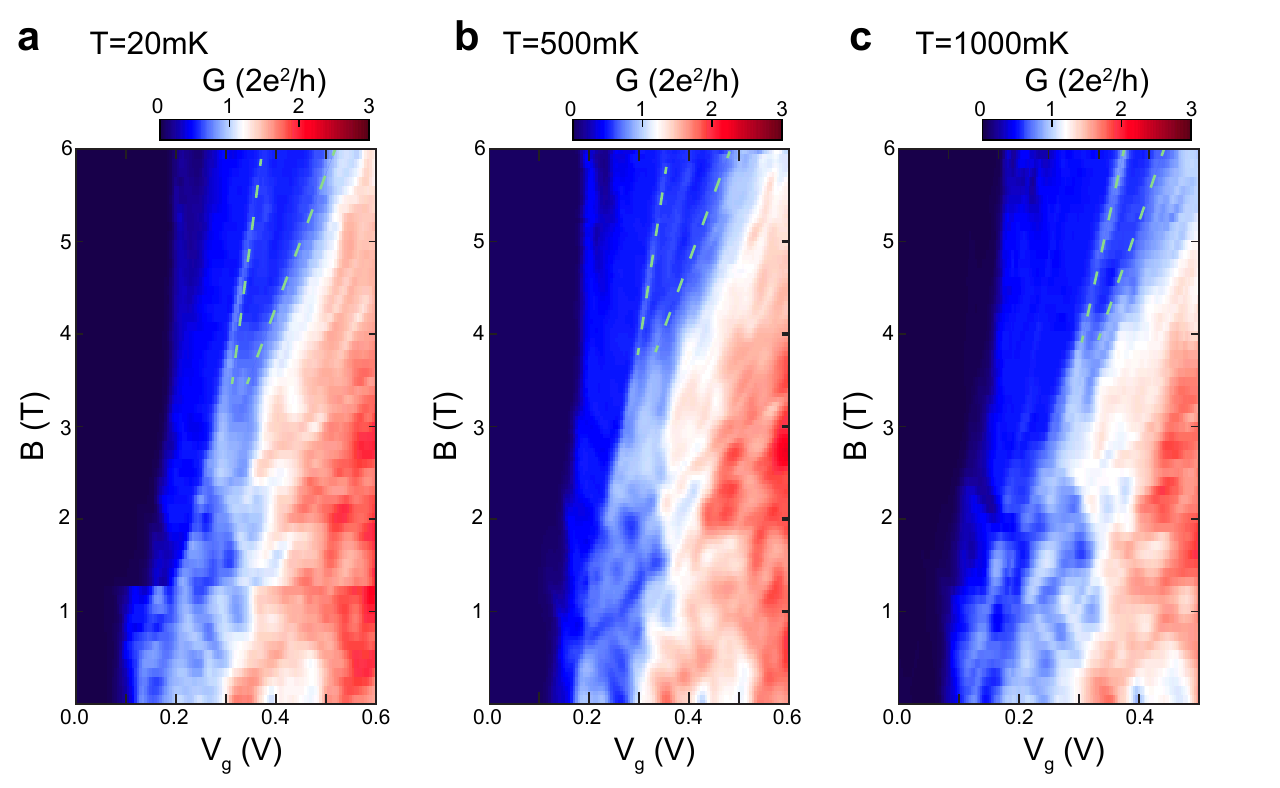}
\caption{ \bf{Temperature dependence of the helical gap. }\normalfont
Measurements of the differential conductance $dI/dV$ ($V_{sd}=\SI{0}{mV}$) as function of magnetic field at 
\bf{a,} \normalfont $\text{T}=\SI{20}{mK}$
\bf{b,} \normalfont $\text{T}=\SI{500}{mK}$
\bf{c,} \normalfont $\text{T}=\SI{1000}{mK}$.
The helical gap (dotted lines) evolves similarly in all three measurements showing that it stays stable at increased temperatures as expected for the energy scale extracted for $\text{E}_{SO}$
}
\label{fig_supp_e2}
\end{figure}

\newpage
\section{Device 2 - Data}
\begin{figure}[!h]
\includegraphics[width=\columnwidth]{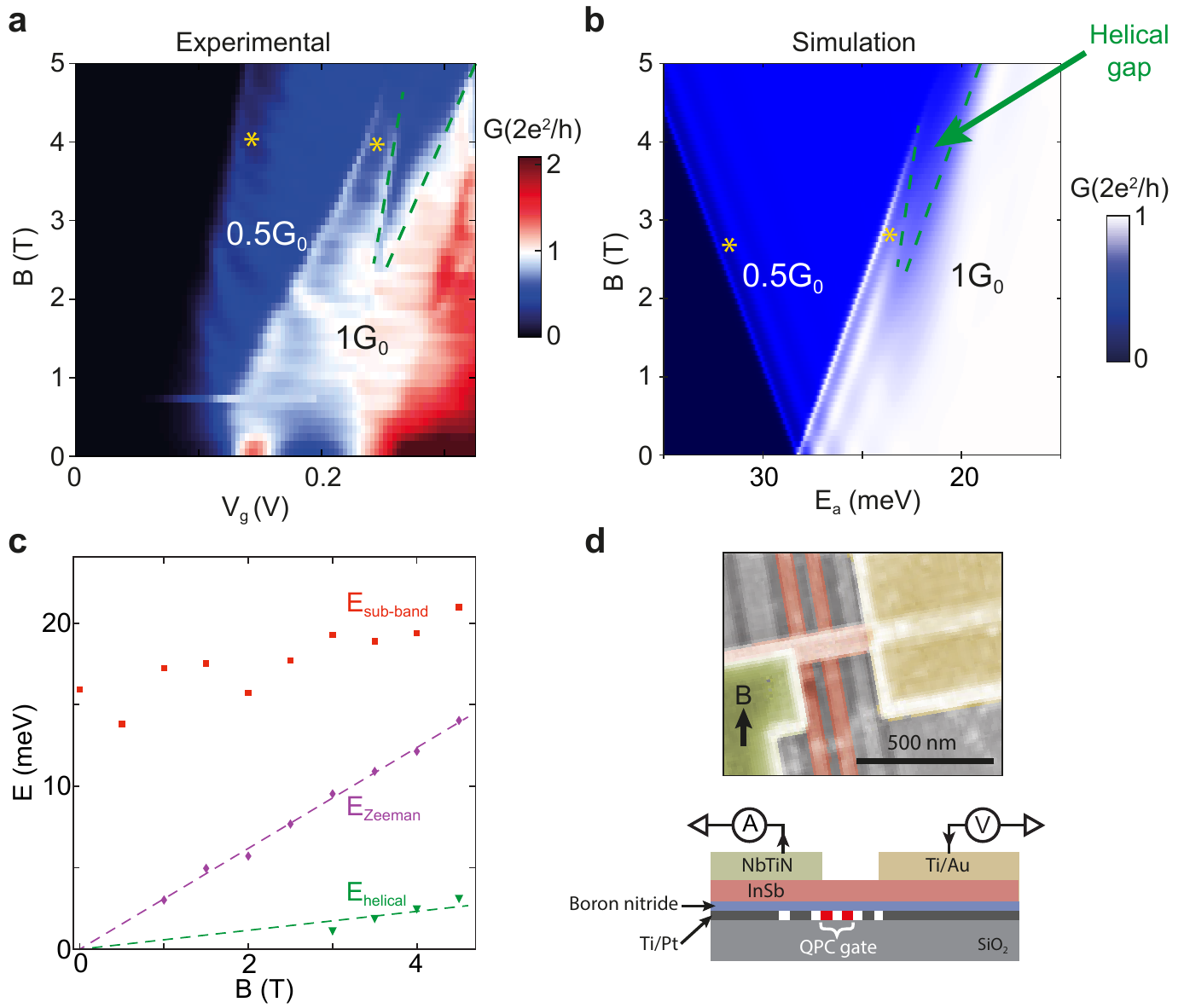}
\caption{\bf{Measurements of a second device. }\normalfont 
\bf{a, }\normalfont Differential conductance $dI/dV$ as function of QPC gate Voltage $\text{V}_g$ and magnetic field B. Around $\text{B} = \SI{2}{T}$ a gap opens in the $1\cdot\text{G}_0$ plateau and increases linearly with magnetic field. At the onset of the $0.5\cdot\text{G}_0$ and the $1\cdot\text{G}_0$ plateaus Fabry-Perot resonances are visible (yellow asterisk). In contrast to the helical gap the width of the resonances stays constant at changing magnetic field.
\bf{b, }\normalfont Numerical simulations of the helical gap with $\theta = \ang{10}$, $g = 53$ and $E_{SO} = \SI{5.6}{meV}$. We use the potential parametrization as for the device discussed in the main text and find a good agreement with the data shown in \bf{a} \normalfont for $\lambda = \SI{40}{nm}$ and $W = \SI{300}{nm}$.
\bf{c, }\normalfont Evolution of the energy levels with magnetic field extracted from the scans shown in Fig \ref{fig_supp_e4}. Dotted lines show fits with intercept fixed at zero and we find a sub-band spacing $\text{E}_{subband}=18\pm \SI{2}{meV}$ and g-factor $g=53\pm 1$. By comparing the slopes of $\text{E}_{Zeeman}\sim \text{E}_Z \cos{\theta}$ and $\text{E}_{helical}\sim \text{E}_Z \sin{\theta}$ we find $\theta = \ang{10} \pm \ang{2}$.
\bf{d, }\normalfont Cross section and false color SEM image of device 2. An InSb nanowire (orange) is contacted by one Ti/Au electrode (yellow) and one NbTiN electrode (green). Two bottom gates (red) are combined to form the QPC constriction. The black arrow indicates the orientation of the applied magnetic field. Measurements are taken at $\SI{20}{mK}$ with the use of standard lock-in technique ($\SI{100}{\micro\volt}$ excitation at \SI{73}{\hertz}).
}
\label{fig_supp_e3}
\end{figure}

\begin{figure}
\includegraphics[width=\columnwidth]{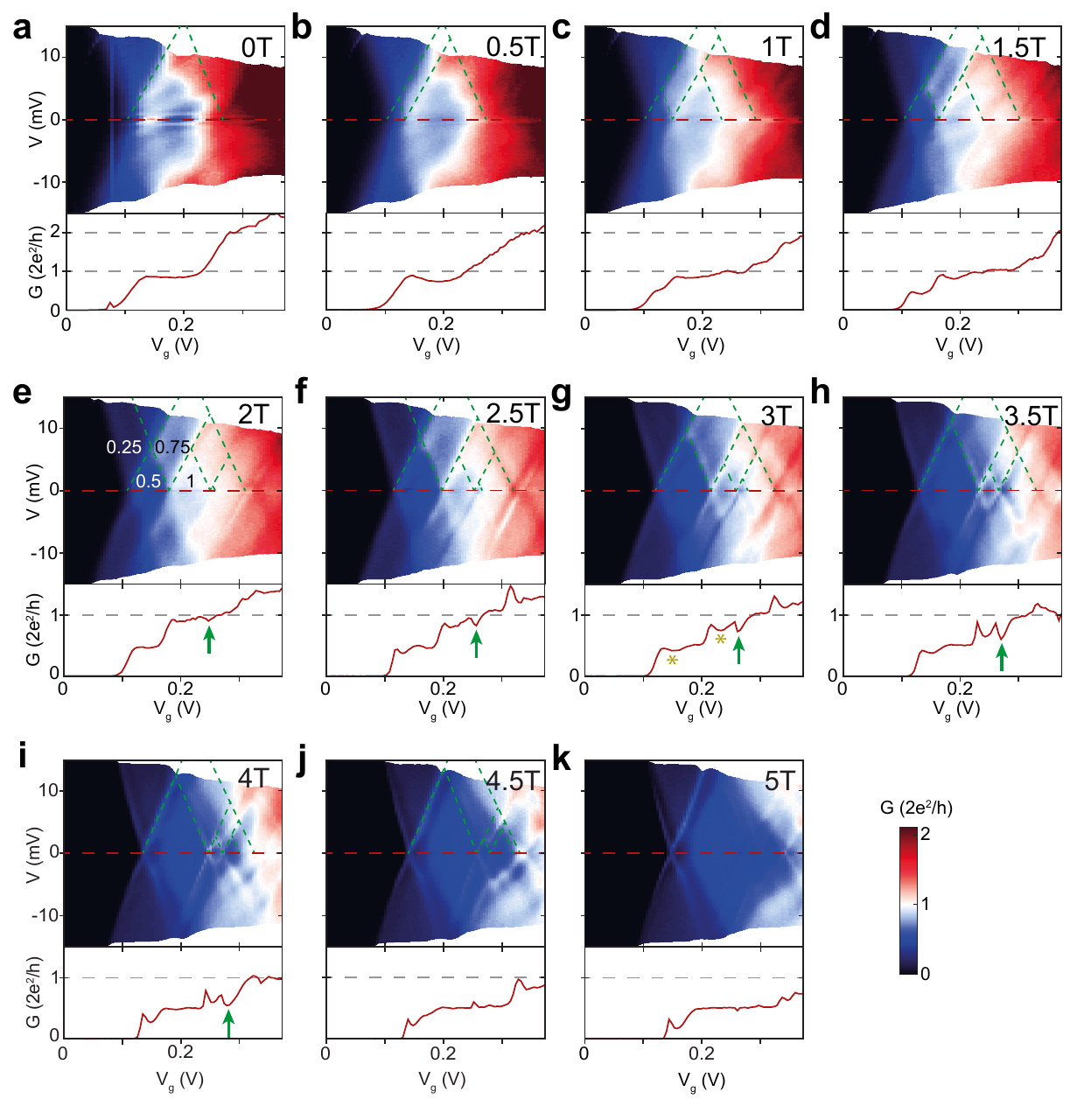}
\caption{ \bf{Voltage bias spectroscopy at increasing magnetic fields. }\normalfont 
\bf{a-k, }\normalfont (Top) Differential conductance $dI/dV$ as a function of QPC gate voltage $\text{V}_g$ and bias voltage $\text{V}_{sd}$. Conductance plateaus show up as diamond shaped region indicated by dashed green lines and can be used to extract the energy spacings shown in Fig \ref{fig_supp_e3}c. Conductance traces in the bottom panels show line cuts taken at $\text{V}_{sd}=\SI{0}{mV}$. Green arrows in \bf{e-i }\normalfont indicate the position of the helical dip. Yellow asterisks in \bf{g }\normalfont indicate conductance dips originating from Fabry-Perot resonances also visible in Fig \ref{fig_supp_e3}a. Numbers in 
\bf{e }\normalfont denote conductance in units of $2e^2/h$.
}
\label{fig_supp_e4}
\end{figure}

\pagebreak

\section{Control devices}

\subsection{QPC length dependence}

\begin{figure}[!h]
\centering
\includegraphics[width=0.8\columnwidth]{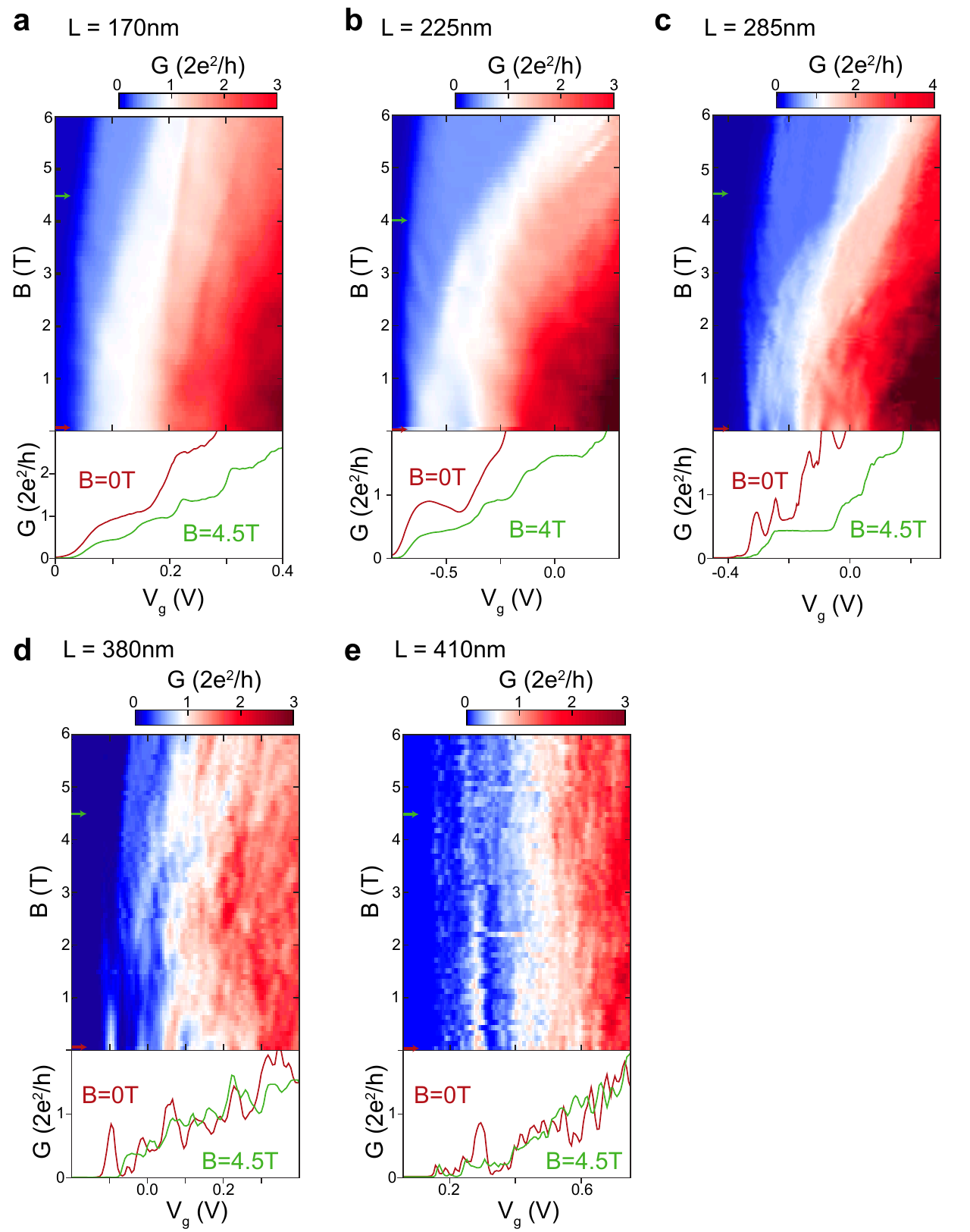}
\caption{ \bf{Length dependence of nanowire QPCs. }\normalfont
Magneto-conductance measurements (at $V_{sd}=\SI{0}{mV}$) of QPCs with increasing length. The contact spacing L is changed in Steps of $\sim \SI{50}{nm}$. 
\bf{a-c, }\normalfont are shorter and \bf{d,e, }\normalfont longer than device 1 ($\text{L} = \SI{325}{nm}$). Line traces at \SI{0}{T} and finite field are added in the bottom panel. The short channel devices \bf{a,b,}\normalfont show well defined and flat plateaus throughout the full magnetic field range. For intermediate channel lengths (\bf{c}\normalfont) resonances start to appear and modify the conductance at low magnetic fields. Long channel devices \bf{d,e, }\normalfont are dominated by backscattering and conductance fluctuations dominate for the full magnetic field range.
}
\label{fig_supp_e5}
\end{figure}

\newpage
\subsection{Simulations - length dependence}
\begin{figure}[!h]
\centering
\includegraphics[width=\columnwidth]{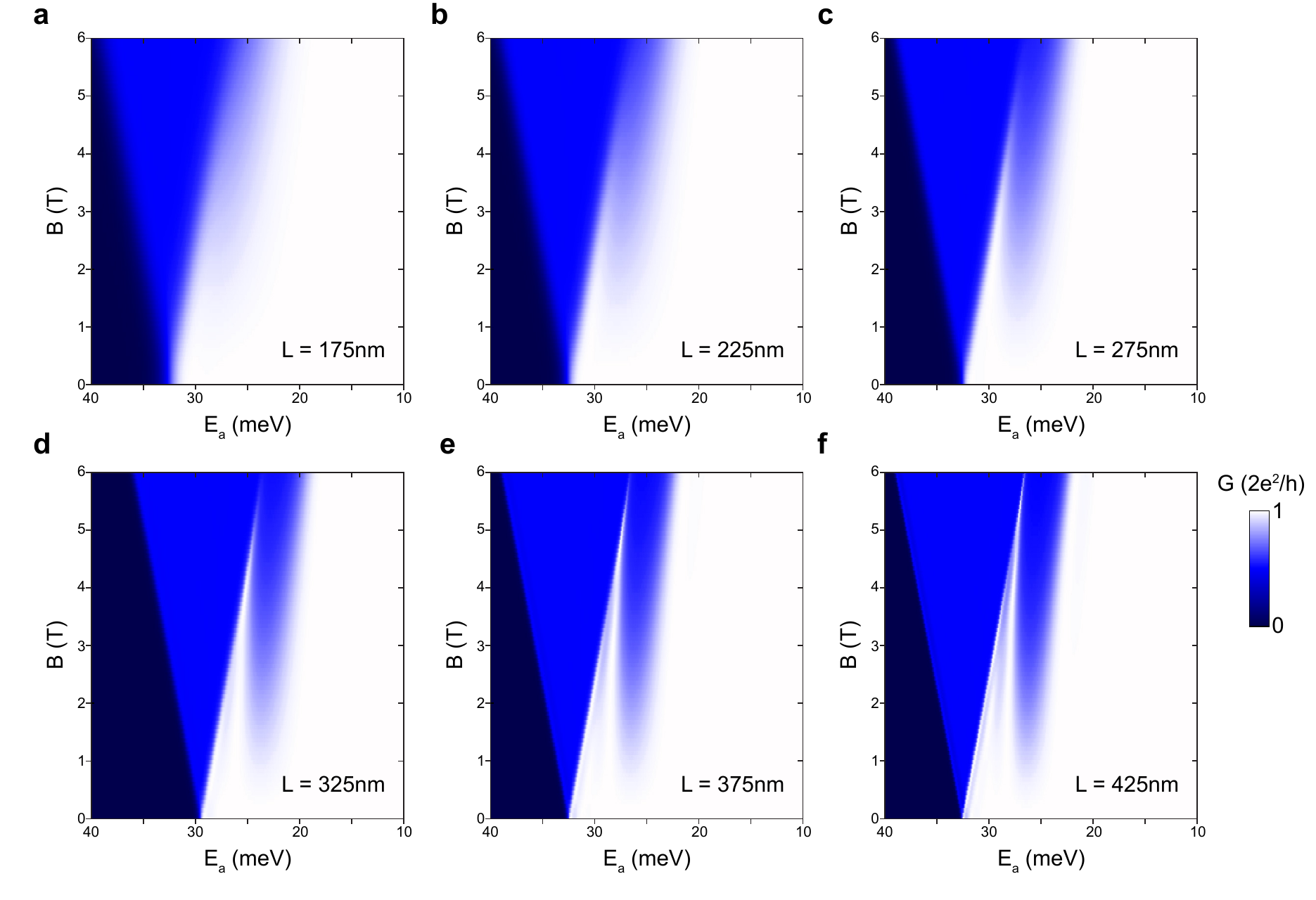}
\caption{ \bf{Simulations for a QPC of different lengths at fixed angle $\theta=\ang{17}$, $l_{SO}=\SI{20}{nm}$. }\normalfont
The contact spacing L is increased in steps of \SI{50}{nm} starting from $L=\SI{175}{nm}$ (\bf{a}\normalfont), up to $L=\SI{425}{nm}$ (\bf{f}\normalfont). The simulations demonstrate the reduced visibility of the helical gap in short devices. At increasing channel length the conductance dip becomes sharper and sets on at lower magnetic fields. A clear reentrant feature can only be seen in \bf{d, e, f, }\normalfont which are at the limit of experimental capabilities (Fig \ref{fig_supp_e5}).}
\label{fig_supp_e6}
\end{figure}

\newpage
\section{Simulations - Angle dependence}
\begin{figure}[!h]
\centering
\includegraphics[width=\columnwidth]{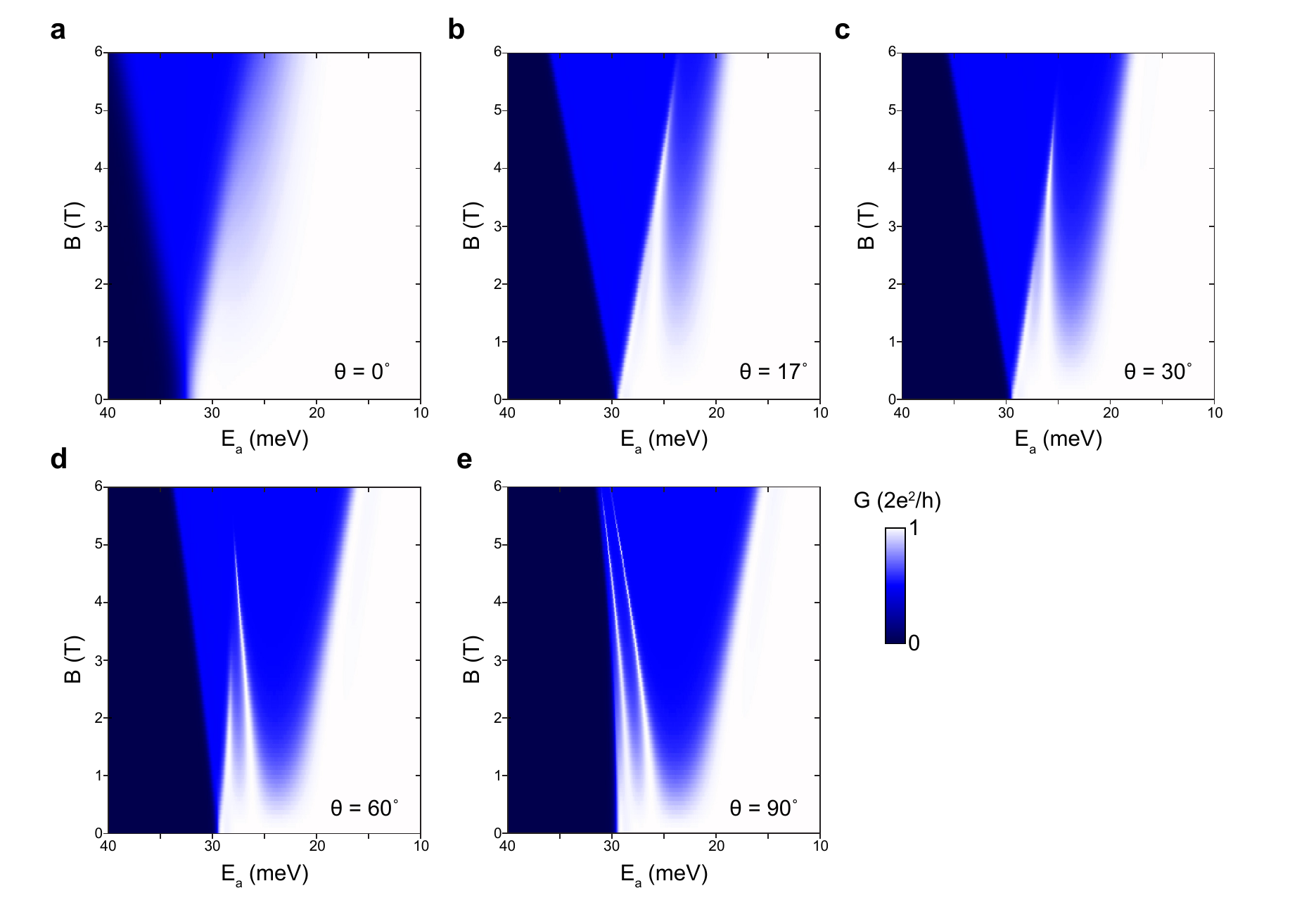}
\caption{ \bf{Simulations of the angle dependence for a QPC with fixed length $L=\SI{325}{nm}$. }\normalfont
$\theta$ is the angle between $\text{B}_{SO}$ and the applied magnetic field as defined in the main text.
\bf{a, }\normalfont For $\theta=\ang{0}$, $\text{B}_{ext}\parallel \text{B}_{SO}$ and the helical gap disappears.
\bf{b-e, } \normalfont at increasing angles $\theta$ the width of the helical gap increases and the width of the inital $0.5\cdot\text{G}_0$ plateau decreases.}
\label{fig_supp_e7}
\end{figure}

\newpage
\section{Simulations - Spin Orbit Strenght}
\begin{figure}[!h]
\centering
\includegraphics[width=\columnwidth]{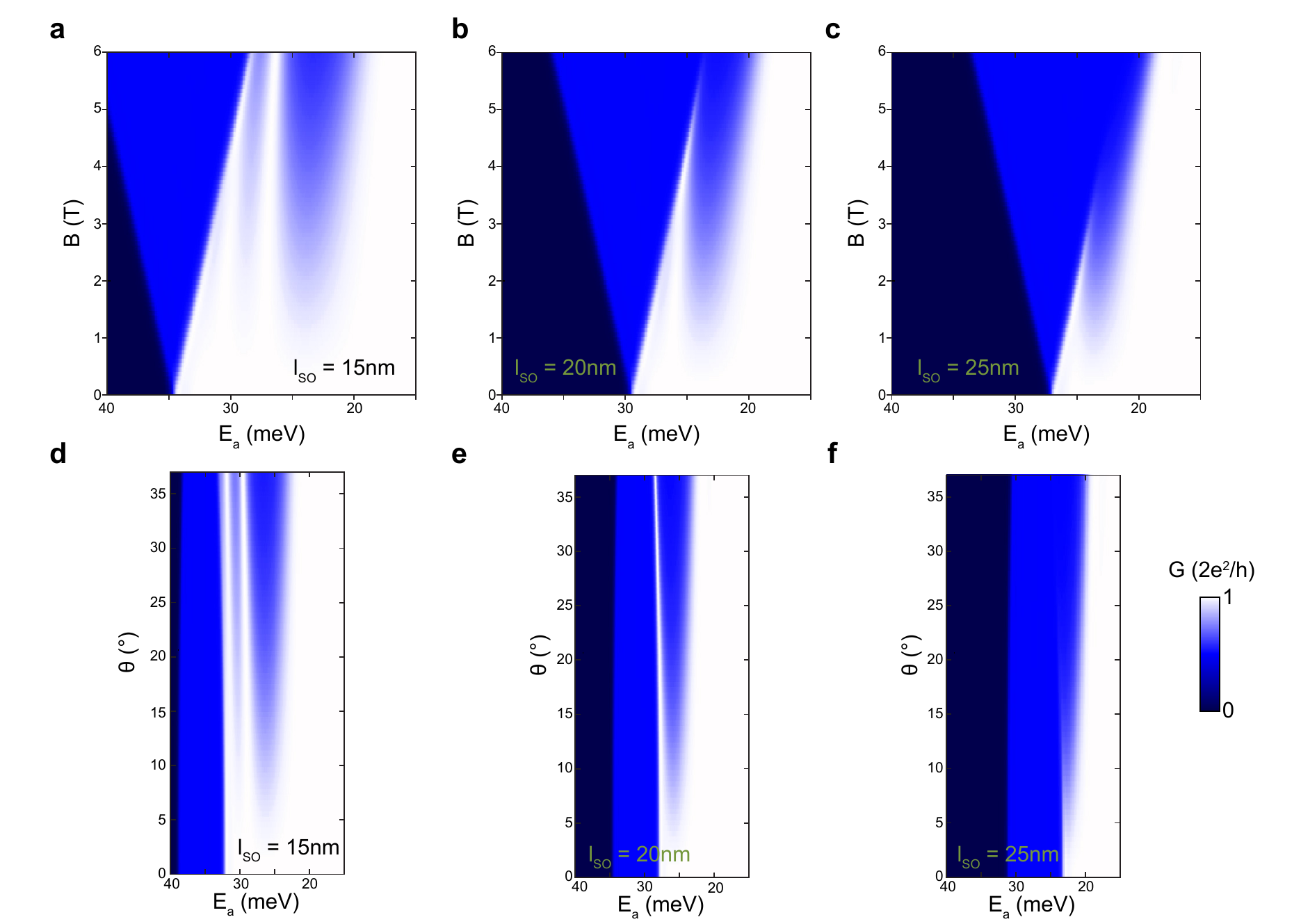}
\caption{ \bf{Simulations of the magnetoconductance for varying $l_{SO}$. }\normalfont
Variations of $\text{l}_{\text{SO}} = 1/k_{\text{SO}}$ strongly influence the visibility of the helical gap in QPC conductance measurements. The simulations for \bf{a, b, c, } \normalfont used identical QPC length $\text{L} = \SI{325}{nm}$ and offset angle $\theta = \ang{17}$
}
\label{fig_supp_e8}
\end{figure}


\begin{thebibliography}{30}

\bibitem{Streda2003} St\v{r}eda, P. \& \v{S}eba, P. Antisymmetric Spin Filtering in One-Dimensional Electron Systems with Uniform Spin-Orbit Coupling. \emph{Phys. Rev. Lett.} \textbf{90}, 256601 (2003)

\bibitem{Pershin2004} Pershin, Y. V., Nesteroff, J. A. \& Privman, Vladimir Effect of spin-orbit interaction and in-plane magnetic field on the conductance of a quasi-one-dimensional system. \emph{Phys. Rev. B} \textbf{69}, 212306 (2004)

\bibitem{Sato2010} Sato, K., Loss, D. \& Tserkovnyak, Y. Cooper-Pair Injection into Quantum Spin Hall Insulators. \emph{Phys. Rev. Lett.} \textbf{105}, 1-4 (2010)

\bibitem{Rashba2016} Shekhter, R. I., Entin-Wohlman, O., Jonson, M. \& Aharony, A. Rashba Splitting of Cooper Pairs. \emph{Phys. Rev. Lett.} \textbf{116}, 1-6 (2016)

\bibitem{alicea2011} Alicea, J., Oreg, Y., Refael, G., Von Oppen, F. \& Fisher, M. P. A. Non-Abelian statistics and topological quantum information processing in 1D wire networks. \emph{Nat. Phys.} \textbf{7}, 412-417 (2011)

\bibitem{Nayak2008} Nayak, C., Simon, S. H., Stern, A., Freedman, M. \& Das Sarma, S. Non-Abelian anyons and topological quantum computation. \emph{Rev. Mod. Phys.} \textbf{80}, 1083--1159 (2008)

\bibitem{Yuval2010} Oreg, Y., Refael, G. \& von Oppen, F, Helical Liquids and Majorana Bound States in Quantum Wires. \emph{Phys. Rev. Lett.} \textbf{105}, 1-4 (2010)

\bibitem{dresselhaus1955} Dresselhaus, G. Spin-Orbit Coupling Effects in Zinc Blende Structures. \emph{Phys. Rev.} \textbf{100}, 580-586 (1955)

\bibitem{rashba2015} Rashba, E. \& Sheka, V. Symmetry of Energy Bands in Crystals of Wurtzite Type II. Symmetry of Bands with Spin-Orbit Interaction Included. \emph{Fiz. Tverd. Tela Collect. Pap.} \textbf{2}, 162--176 (1959)

\bibitem{winkler2003} Winkler, R. \emph{Spin-orbit Coupling Effects in Two-Dimensional Electron and Hole Systems}. (Springer-Verlag Berlin Heidelberg, 2003)
\emph{} \textbf{}, ()

\bibitem{Koenig2007} K{\"o}nig, M. \emph{et al}.  Quantum Spin Hall Insulator State in HgTe Quantum Wells. \emph{Science (80-. )}. \textbf{318}, 766–771 (2007)

\bibitem{nowack2013} Nowack, K. C. \emph{et al}. Imaging currents in HgTe quantum wells in the quantum spin Hall regime. \emph{Nat. Mater.} 12, 787–791 (2013)

\bibitem{quay2010} Quay, C. H. L., Hughes, T. L., Sulpizio, J. A., Pfeiffer, L.N. , Baldwin, K. W., West, K.W., Goldhaber-Gordon, D. \& De Picciotto, R. Observation of a one-dimensional spin–orbit gap in a quantum wire. \emph{Nat. Phys.} \textbf{6}, 336–339 (2010).

\bibitem{klinovaja2011} Klinovaja, J., Schmidt, M. J., Braunecker, B. \& Loss, D. Helical modes in carbon nanotubes generated by strong electric fields. \emph{Phys. Rev. Lett.} \textbf{106}, 1–4 (2011).

\bibitem{klinovaja2013topological} Klinovaja, J., Stano, P., Yazdani, A. \& Loss, D. Topological superconductivity and Majorana fermions in RKKY systems. \emph{Phys. Rev. Lett.} \textbf{111}, 1–5 (2013).

\bibitem{mourik2012signatures} Mourik, V., Zuo, K.,Frolov, S. M., Plissard, S.R., Bakkers, E. P. A. M. \& Kouwenhoven, L. P. Signatures of Majorana Fermions in hybrid superconductor-semiconductor nanowire devices. \emph{Science (80-. )}. \textbf{336}, 1003 (2012).

\bibitem{albrecht2016exponential} Albrecht, S. M. \emph{et al}. Exponential Protection of Zero Modes in Majorana Islands. \emph{Nature} \textbf{531}, 206–209 (2016).

\bibitem{Rainis2014} Rainis, D. \& Loss, D. Conductance behavior in nanowires with spin-orbit interaction: A numerical study. \emph{Phys. Rev. B} \textbf{90}, 1–9 (2014).

\bibitem{kammhuber2016conductance} Kammhuber, J. \emph{et al}. Conductance Quantization at Zero Magnetic Field in InSb Nanowires. \emph{Nano Lett.} \textbf{16}, 3482–3486 (2016).

\bibitem{heedt2016ballistic} Heedt, S., Prost, W., Schubert, J., Grützmacher, D. \& Schäpers, T. Ballistic Transport and Exchange Interaction in InAs Nanowire Quantum Point Contacts. \emph{Nano Lett.} \textbf{16}, 3116–3123 (2016).

\bibitem{van2012quantized} Van Weperen, I., Plissard, S. R., Bakkers, E. P. A. M., Frolov, S. M., Kouwenhoven, L. P. Quantized conductance in an InSb nanowire. \emph{Nano Lett.} \textbf{13}, 387–391 (2013).

\bibitem{nadj2012spectroscopy} Nadj-Perge, S. \emph{et al}. Spectroscopy of spin-orbit quantum bits in indium antimonide nanowires. \emph{Phys. Rev. Lett.} \textbf{108}, 1–5 (2012).

\bibitem{van2015spin} Van Weperen, I. \emph{et al}. Spin-orbit interaction in InSb nanowires. \emph{Phys. Rev. B} \textbf{91}, 1–17 (2015).

\bibitem{cayao2015sns} Cayao, J., Prada, E. , San-Jose, P. \& Aguado, R. SNS junctions in nanowires with spin-orbit coupling: Role of confinement and helicity on the subgap spectrum. \emph{Phys. Rev. B} \textbf{91}, 24514 (2015).

\bibitem{heyder2015relation} Heyder, J. \emph{et al}. Relation between the 0.7 anomaly and the Kondo effect: Geometric crossover between a quantum point contact and a Kondo quantum dot. \emph{Phys. Rev. B} \textbf{92}, (2015).

\bibitem{Goulko2014} Goulko, O., Bauer, F., Heyder, J. \& Von Delft, J. Effect of spin-orbit interactions on the 0.7 anomaly in quantum point contacts. \emph{Phys. Rev. Lett.} \textbf{113}, 1–5 (2014).

\bibitem{sau2012experimental} Sau, J. D., Tewari, S. \& Das Sarma, S. Experimental and materials considerations for the topological superconducting state in electron- and hole-doped semiconductors: Searching for non-Abelian Majorana modes in 1D nanowires and 2D heterostructures. \emph{Phys. Rev. B} \textbf{85}, 1–11 (2012).

\bibitem{plissard2012insb} Plissard, S. R. \emph{et al}. From InSb nanowires to nanocubes: Looking for the sweet spot. \emph{Nano Lett.} \textbf{12}, 1794–1798 (2012).

\bibitem{logg2012automated} Logg, A., Mardal, K.-A., Wells, G. N. \emph{et al}. Automated Solution of Differential Equations by the Finite Element Method, \emph{Springer}, 2012.

\bibitem{groth2014kwant} C. W. Groth, C. W., Wimmer, M., Akhmerov, A. R., Waintal, X., Kwant: a software package for quantum transport, \emph{New J. Phys.} \textbf{16}, 063065 (2014)

\end{thebibliography}

\begin{thebibliography}{6}

\bibitem{supp1}	D. Rainis and D. Loss Conductance behavior in nanowires with spin-orbit interaction: A numerical study. \emph{Phys. Rev. B} \textbf{90}, 235415 (2014).
\bibitem{supp2}	N. March The Thomas-Fermi approximation in quantum mechanics. \emph{Advances in Physics} \textbf{6}, 1 (1957)
\bibitem{supp3}	A. Logg, K.-A. Mardal, G. N. Wells, et al., \emph{Automated Solution of Differential Equations by the Finite Element Method} (Springer, 2012).
\bibitem{supp4} V. Eyert, \emph{Journal of Computational Physics} \textbf{124}, 271 (1996).
\bibitem{supp5} C. W. Groth, M. Wimmer, A. R. Akhmerov, and X. Waintal, \emph{New Journal of Physics} \textbf{16}, 063065 (2014) 
\bibitem{supp6} Y. V. Pershin, J. A. Nesteroff, and V. Privman, \emph{Phys. Rev. B} \textbf{69}, 121306 (2004)

\end{thebibliography}
\end{document}